\documentclass[a4paper,11pt]{article}

\usepackage{pos}
\usepackage{caption}
\usepackage{cellspace}
\usepackage{float}
\usepackage{lineno}
\usepackage{placeins}

\title{To cut or not to cut: Data-quality evaluation for the ASTRI Mini-Array}
\ShortTitle{Data-quality evaluation for the ASTRI Mini-Array}

\author[a]{E. Molina}
\author*[b]{L. Foffano}
\author[c]{S. Crestan}
\author[d]{S. Iovenitti}
\author[e,f]{S. Lombardi}
\author[e,f]{F. Lucarelli}
\author[g]{T. Mineo,}
\onbehalf{for the ASTRI Project\footnote[2]{\url{http://www.astri.inaf.it/en/library/}}}
\renewcommand{\printHeadAuthors}{E. Molina}

\affiliation[a]{Instituto de Astrof\'isica de Canarias and Dpto. de Astrof\'isica, Universidad de La Laguna, \\
C/ V\'ia L\'actea s/n, E-38205 La Laguna, Tenerife, Spain}

\affiliation[b]{INAF - IAPS, \\
via del Fosso del Cavaliere 100, I-00133 Roma, Italy}

\affiliation[c]{INAF - IASF Milano, \\
Via Alfonso Corti 12, 20133 Milano, Italy}

\affiliation[d]{INAF - Osservatorio astronomico di Brera, \\
Via E. Bianchi 46, 23807, Merate, Italy}

\affiliation[e]{INAF - OAR, \\
Via Frascati 33, I-00078 Monte Porzio Catone (Roma), Italy}

\affiliation[f]{ASI, Space Science Data Center, \\
Via del Politecnico s.n.c., I-00133 Roma, Italy}

\affiliation[g]{INAF - IASF Palermo, \\
Via U. La Malfa 153, 90146 Palermo, Italy}

\emailAdd{emolina@iac.es}
\emailAdd{luca.foffano@inaf.it}

\abstract{The ASTRI Mini-Array consists of nine Cherenkov telescopes under construction at the Observatorio del Teide, with one of them already performing observations since November 2024. Given the complexity of the analysis of gamma-ray data acquired with Cherenkov telescopes, a proper evaluation of the quality of these data is important to avoid a negative impact on the high-level scientific products. There are several factors that can contribute to reducing the quality of the observations, such as clouds, high humidity or a high dust concentration in the air, among others. In order to take all these factors into account and evaluate their impact on the data, a quality check pipeline has been developed for ASTRI. This contribution describes the pipeline, from its inputs at different data levels to the production of outputs in the form of good-quality lists and diagnostic plots. We also show how setting preliminary quality cuts on the Crab Nebula data taken with the first ASTRI telescope can significantly improve the source detection.}

\ConferenceLogo{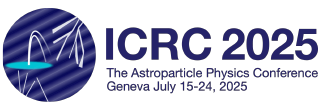}

\FullConference{39th International Cosmic Ray Conference (ICRC2025) \\
                15–24 July 2025                                     \\
                Geneva, Switzerland}

\begin{document}
\maketitle

\section{Introduction}

The ASTRI Mini-Array \citep{astri22_general} is an array of nine imaging atmospheric Cherenkov telescopes located at the Observatorio del Teide ($28^\circ 18^\prime 04^{\prime\prime}$~N, $16^\circ 30^\prime 38^{\prime\prime}$~W, 2390~m above sea level) in the island of Tenerife, Spain. The array is currently under construction, with one telescope (ASTRI-1) being already in a commissioning phase and taking data since November 2024, and the rest expected to become operational over the second half of 2025 and 2026 \citep{pareschi25}. Each telescope follows a modified Schwarzschild-Couder optical configuration \citep{vassiliev07,sironi17}, with a primary and a secondary mirror with diameters of 4.3~m and 1.8~m, respectively. The camera is made of silicon photomultipliers and has 2368 pixels, covering a field of view (FoV) of $10.5^\circ$ with an angular pixel size of $0.19^\circ$. According to the simulations, the ASTRI Mini-Array is sensitive to gamma rays with energies between 0.3 and 200~TeV, with an energy resolution of 12\% and an angular resolution of $0.05^\circ$ at 10~TeV. The performance of the array remains approximately constant up to an angular distance of $3^\circ$ from the camera centre. This fact, together with its large FoV, makes ASTRI a great instrument to study extended sources at TeV energies and above. A more detailed description of the array performance can be found in \cite{astri22_science}.

Together with the nine Cherenkov telescopes responsible for taking scientific data, there are four main auxiliary systems that provide real-time information on the environmental conditions: two weather stations, which measure different weather parameters like humidity, temperature and pressure; the sky quality meter (SQM), which measures the sky brightness in the same direction as the telescopes are pointing to; and the all-sky camera, which provides information on the cloudiness (fraction of the sky covered by clouds). In the future, a LIDAR will also be installed in order to monitor the atmospheric transmission within the FoV. Moreover, the Cherenkov camera also records so-called variance data \citep{segreto19}, which provide additional information on the number of stars in FoV and on the effect of the night sky background on the camera pixels. These quantities can be used, among other things, to characterize the environmental conditions or the telescope mispointing.

In this contribution, we present how a preliminary evaluation of the scientific data quality from the ASTRI-1 telescope is performed using an end-to-end pipeline. The pipeline uses the scientific data themselves, as well as the information provided by the auxiliary systems and the variance data. Section~\ref{sec:observations} describes the ASTRI-1 observations used for the data-quality study. Section~\ref{sec:pipeline} provides a brief technical description of the pipeline and of its main processing steps, including the data selection. A preliminary example of the improved scientific results based on the quality cuts applied with this pipeline is shown in Section~\ref{sec:results}, and some concluding remarks are given in Sec.~\ref{sec:conclusions}.

\section{Description of the observations}\label{sec:observations}

This quality study is based on 200.8 hours of Crab Nebula observations performed between November 2024 and February 2025. The data span four distinct observational periods separated by several days around the full Moon, when the telescope was not operating. A summary of these periods is provided in Table~\ref{tab:observations}.

In order to better estimate the background accounting for the non-uniform sensitivity of the camera, all the observations were done in wobble mode \citep{fomin94}, in which the source is located at an offset angle with respect to the camera centre. Four different source positions were normally used for a given offset, two along the right ascension axis and two along the declination axis. Each observation run has the source located in one of these four positions, and consecutive runs alternate between different positions until the cycle is completed and restarted. Typically, a run lasts around 30~min, so the whole cycle takes $\sim2$~h to complete.

Since the ASTRI-1 telescope is in a commissioning phase, the observations were done using varying offset angles --ranging from 0.5$^\circ$ to 4.5$^\circ$-- in order to explore the telescope capabilities in different conditions. For the purpose of this work, we only considered observations taken with a 0.5$^\circ$ wobble offset. Moreover, in order to minimize the variation of several quantities with the zenith angle, we restrict our study to the data obtained when the source was at a zenith angle below $30^\circ$. The corresponding times for each period that result from applying these conditions are stated in the last column of Table~\ref{tab:observations}, for a total of 52.3~h.

\begin{table}
    \captionsetup{width=0.7\linewidth, skip=-2pt}
    \caption{Observing times of the Crab Nebula with ASTRI-1 for each of the periods defined by the start and end dates, together with the total observing time. The last column shows the times for the data used in this work, with an offset angle of 0.5$^\circ$ and zenith angles below 30$^\circ$ (see text).}
    \label{tab:observations}
    \begin{center}
    \begin{tabular}{ScScScSc}
        \hline
        \hline
        Start date  & End date      & Observing time (h)    & Selected time (h) \\
        \hline
        2024-11-22  & 2024-12-09    & 71.2                  & 20.9              \\
        \hline
        2024-12-27  & 2025-01-08    & 56.0                  & 5.9               \\
        \hline
        2025-01-20  & 2025-02-03    & 46.6                  & 11.9              \\
        \hline
        2025-02-18  & 2025-02-27    & 27.0                  & 13.6              \\
        \hline
        \multicolumn{2}{c}{Total}   & 200.8                 & 52.3              \\
        \hline
    \end{tabular}
    \end{center}
\end{table}

\section{The ASTRI data-quality check pipeline}\label{sec:pipeline}

The quality evaluation of the ASTRI-1 data is done within a data-quality check (DQC) pipeline that takes care of all the necessary steps to perform such an evaluation. The pipeline operates with FITS input files at different data levels and obtained by the Cherenkov camera and the auxiliary systems. Files are produced on a run basis, i.e., every observing run has associated its own set of scientific and auxiliary files. Regarding the scientific runs, the pipeline uses the raw information recorded by the camera, as well as some quantities computed at later stages of the data processing, such as the image (Hillas) parameters \citep{hillas85} and the reconstructed energy, arrival direction and \textit{gammaness} of each event (see \cite{astri22_science, crestan25} for more information on the computation of these quantities as part of the global analysis chain). Since the DQC analysis uses processed data that become available only after the scientific observations are completed, the pipeline is typically executed the following morning, once the ASTRI data processing and reduction have been finalized.

\begin{figure}
    \centering
    \includegraphics[width=0.7\linewidth]{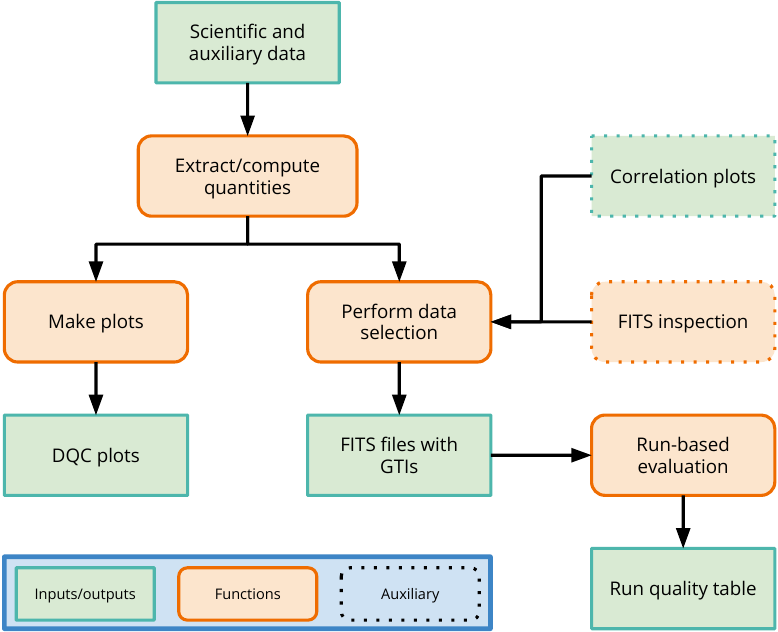}
    \captionsetup{width=0.8\linewidth, skip=10pt}
    \caption{Schematic representation of the the ASTRI DQC pipeline. Orange boxes represent the steps performed by the pipeline, while green boxes are the inputs and outputs of each step. Dotted boundary lines are used for auxiliary functions and inputs not belonging to the automatic pipeline itself, but that are employed for the data-quality selection step.}
    \label{fig:pipeline}
\end{figure}

\begin{figure}
    \centering
    \includegraphics[width=\linewidth]{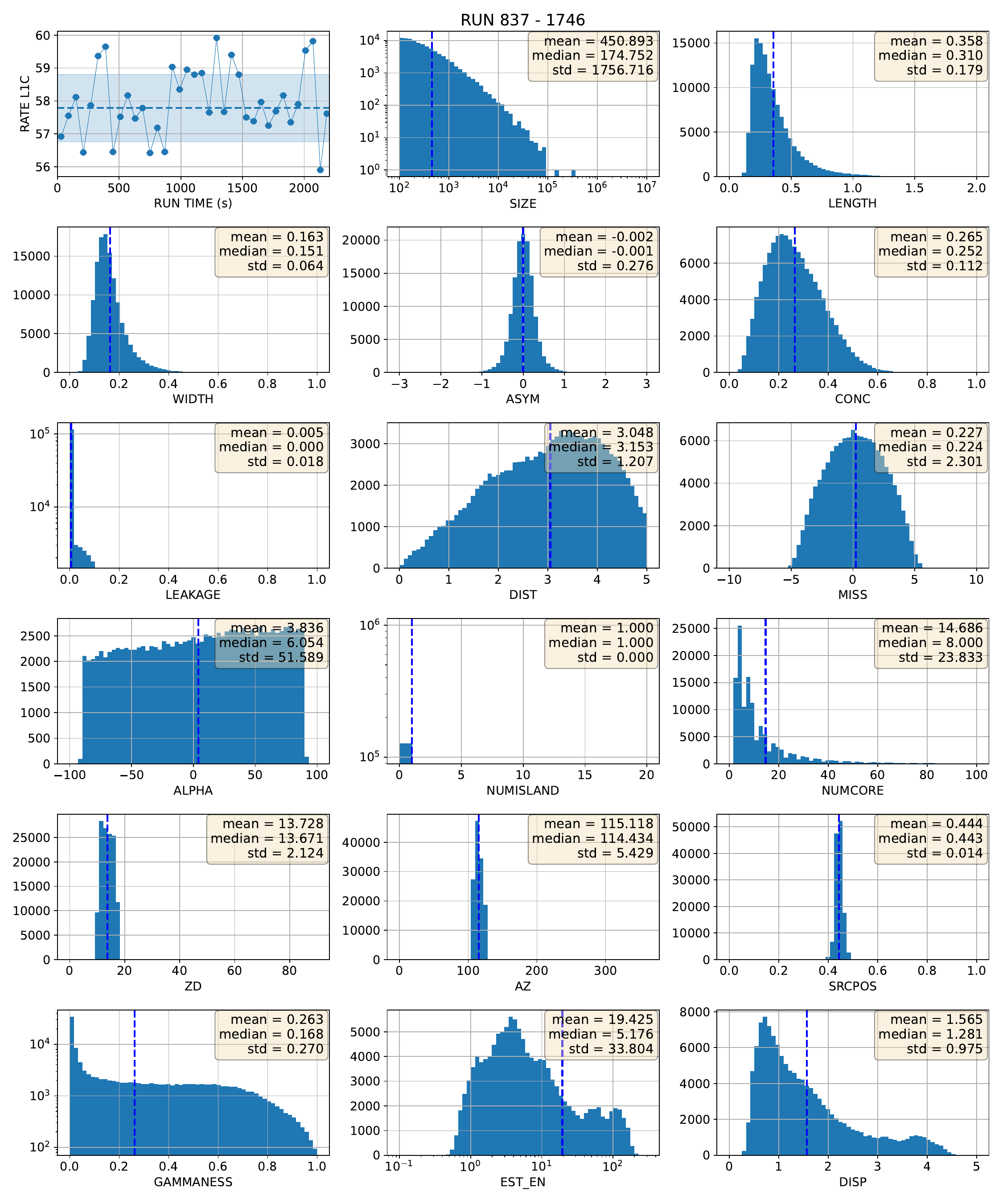}
    \captionsetup{width=\linewidth, skip=10pt}
    \caption{Representative example of some of the plots produced by the DQC pipeline for a given observing run. The intra-run evolution of the event rate is shown in 1-min time bins, together with the histograms of several quantities.}
    \label{fig:dqc_plots}
\end{figure}

The DQC pipeline runs in a computing cluster physically located at the Osservatorio Astronomico di Roma, in Italy. Several configuration parameters are set through a TOML file. Among them, we include the date range of the data to evaluate and the set of quality cuts that will be used in the selection procedure. Once executed, the pipeline automatically performs a series of sequential steps that end up in the production of different outputs ready to use in later stages of the ASTRI data analysis. Specifically, three kinds of outputs are generated: a list of good-quality runs, a list of good time intervals (GTIs) within single runs, and diagnostic plots for human inspection (see examples of the latter in Fig.~\ref{fig:dqc_plots}). A schematic representation of the DQC pipeline execution is provided in Fig.~\ref{fig:pipeline}.

\section{Preliminary results}\label{sec:results}

Here we present preliminary results from the analysis of the Crab Nebula data described in Sec.~\ref{sec:observations}, highlighting the improved outcomes obtained by applying the quality cuts defined in this work. These results were obtained by using the 52.3~h of data listed in Table~\ref{tab:observations}. The definition of the proper quantities that characterize the quality of the data and the study of their effect on the high-level scientific results is an ongoing work. Currently, this data-quality selection is based on an empirical approach applied to the Crab Nebula observations, and will evolve in the next months as more data from different FoVs and telescopes become available. Therefore, the results presented here should be taken as very preliminary.

In order to select the parameters that best define the quality of the data, we studied the correlations between different quantities. In particular, we used the event rate of the reduced (and cleaned) data as a reasonably good indicator of the data-taking conditions, and checked how other quantities changed with it. We recall that we limit our study to zenith angles below $30^\circ$, where the change of rate with the pointing position is small compared to other effects. The result of this study was the selection of three different cuts that improve the quality of the data, each of them taking care of different aspects. These cuts are always considered at a run level, leaving the GTI-based data selection as future work:

\begin{enumerate}

\item \textbf{Median background from the variance data below 6 squared photo-electron}. This cut ensures the usage of data taken under dark conditions, removing observations performed during moon time or twilight.

\item \textbf{Number of detected stars within the FoV higher than 20}. For Crab Nebula observations, this corresponds to more than $\sim 75$\% of the stars detected in optimal conditions. Lower values of the number of stars indicate poor weather conditions, mainly owing to the presence of clouds or high humidity.

\item \textbf{Relative standard deviation of the cleaned rate below 10\%}. This condition is set in order to get rid of runs with high internal variability due to different reasons (clouds passing by, moon rising or setting, electronic problems,...).

\end{enumerate}

\begin{figure}
    \centering
    \includegraphics[width=0.49\linewidth]{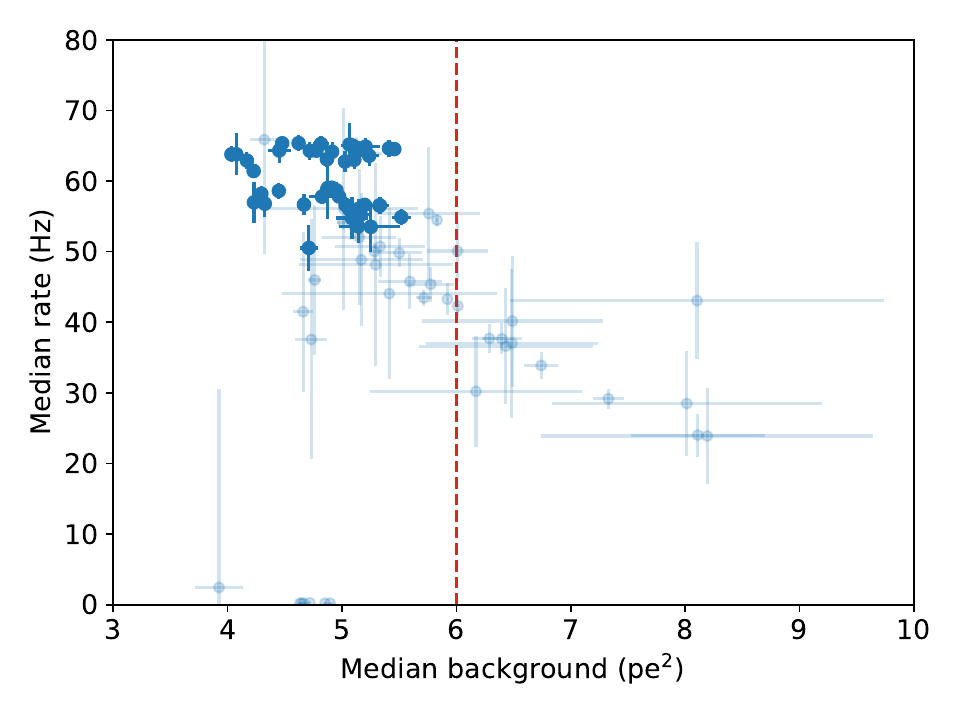}
    \includegraphics[width=0.49\linewidth]{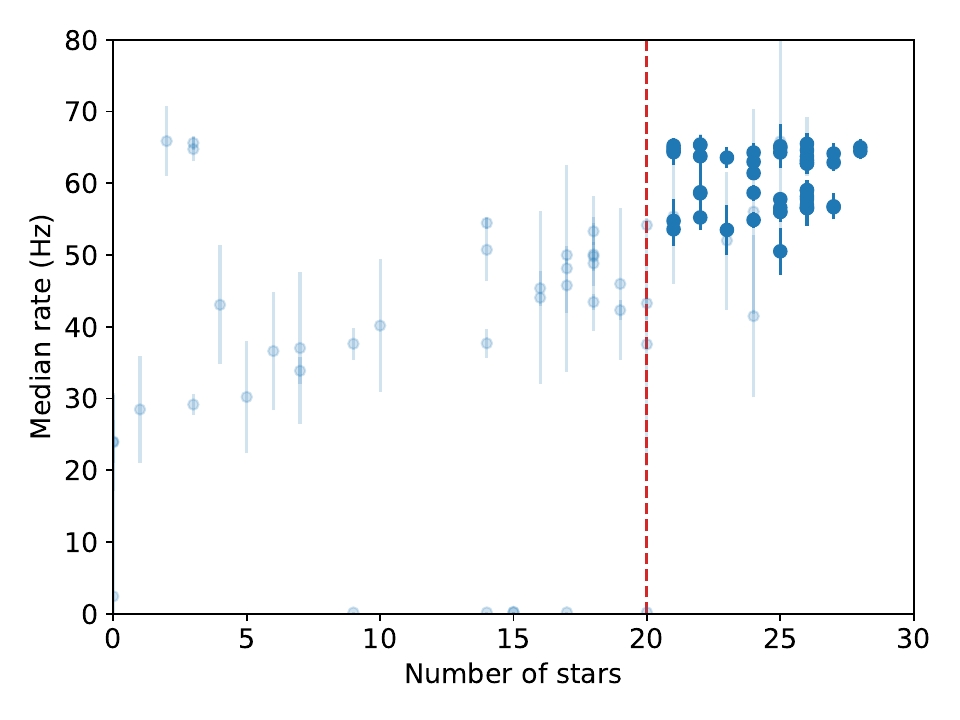}
    \captionsetup{width=\linewidth, skip=10pt}
    \caption{Dependency between the rate of the cleaned scientific data runs with the background (left) and number of detected stars (right) computed from variance data. The error bars represent the standard deviation of each quantity within the run. The highlighted points are those fulfilling the 3 quality cuts described in Sec.~\ref{sec:results}. The dashed red lines represent the cuts in the background and number of stars in the left and right panels, respectively.}
    \label{fig:rate_correlations}
\end{figure}

Figure~\ref{fig:rate_correlations} shows how the median run rate of the cleaned data varies with the background and the number of detected stars computed from the variance data. As seen from the highlighted points in the figure, setting the cuts described above ensures a stable rate above 50~Hz, which is a good indicator regarding the quality of the observations. In order to confirm the adequate data-taking conditions of the selected runs, we also performed further checks (not shown here for simplicity) by visually inspecting the evolution of some quantities in the (intra-run) diagnostic plots produced by the DQC pipeline.


\begin{figure}
    \centering
    \includegraphics[width=0.49\linewidth]{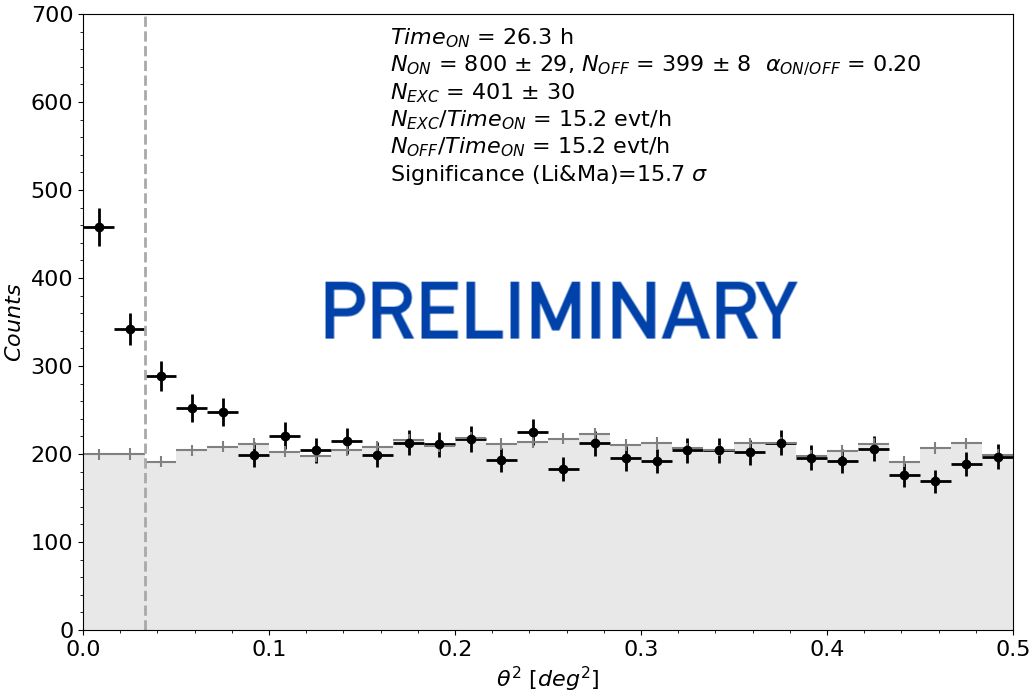}
    \includegraphics[width=0.49\linewidth]{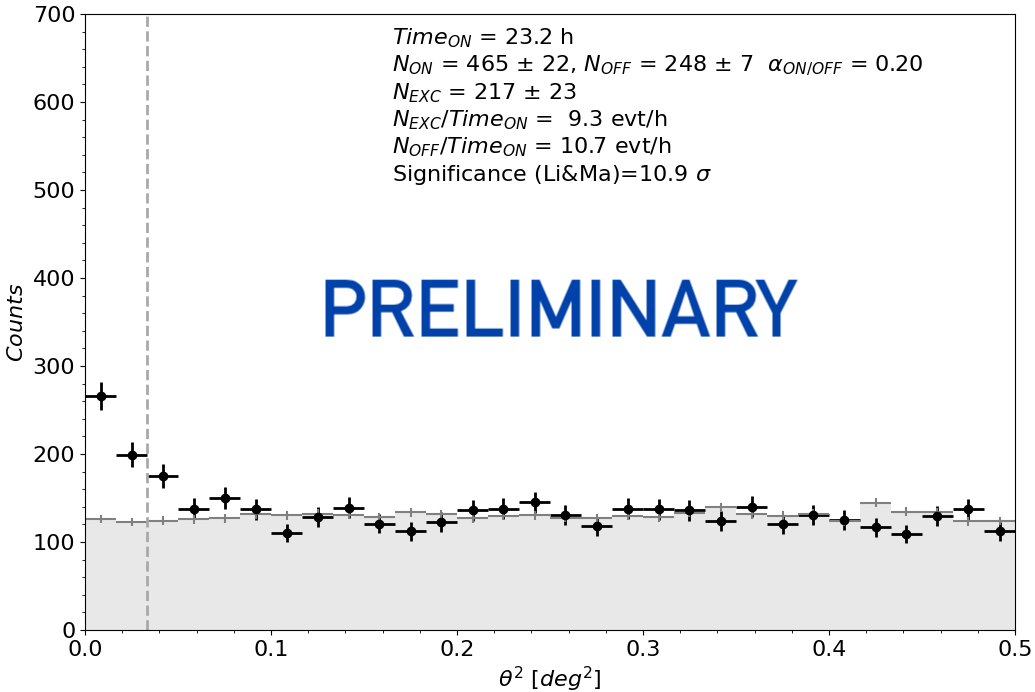}
    \captionsetup{width=\linewidth, skip=10pt}
    \caption{Detection $\theta^2$ plots for the Crab Nebula data passing the quality cuts (left) and those rejected by the data-quality selection (right). The black crosses represent the number of (ON) events coming from a circular region centred around the source, while the grey area are the background (OFF) events computed from a symmetric region with respect to the camera centre. The vertical dashed line at 0.033~deg$^2$ represents the limit of the region from which the number of ON and OFF events are taken in order to compute the detection significance of the source.}
    \label{fig:theta2_plots}
\end{figure}

As a preliminary evaluation of the performance of the quality study,  we produced $\theta^2$ detection plots of the Crab Nebula for both the runs selected (Fig~\ref{fig:theta2_plots}, left panel) or rejected (Fig~\ref{fig:theta2_plots}, right panel) by the three cuts above. We can clearly see an improvement of the Li\&Ma detection significance \citep{li83} of the good-quality data, $15.7\sigma$, with respect to the bad-quality ones, $10.9\sigma$, with similar observing times of 26.3~h and 23.2~h, respectively\footnote{The 2.8~h discrepancy between the sum of these times and the 52.3~h reported in Sec.\ref{sec:observations} is explained by the rejection of some individual events by the data reduction chain, before producing the $\theta^2$ plots.}. We can further evaluate this improvement by computing the rate of excess events in each case, which is $15.2 \pm 1.1$~events/h for the good-quality data, and $9.3 \pm 1.0$~events/h for the bad-quality data, resulting in an enhancement of 63\%. Given that both $\theta^2$ plots were produced using the same data analysis configuration, we can attribute the observed improvement to the applied quality cuts.

\section{Conclusions}\label{sec:conclusions}

The data-quality evaluation of the ASTRI Mini-Array is done through an end-to-end pipeline that takes care of performing a series of steps required to conduct such evaluation and produce different outputs to be used at further steps of the ASTRI data analysis. In this work, the DQC pipeline was applied to a subset of the Crab Nebula data obtained during the commissioning phase of the ASTRI-1 telescope, for a total of 52.3~h. The results of the data-quality selection, even if indicative and based on a preliminary choice of the quality parameters, already suggest a significant improvement in the signal obtained for the good-quality data with respect to the bad-quality ones, pointing at the importance of the data selection as part of the overall analysis of ASTRI data. In this direction, further work is being performed in order to refine and improve the selection procedure.

\acknowledgments
This work was conducted in the context of the ASTRI Project. We gratefully acknowledge support from the people, agencies, and organisations listed here: \url{http://www.astri.inaf.it/en/library/}. EM acknowledges support from the Agencia Estatal de Investigación from the Spanish Ministerio de Ciencia, Innovación y Universidades (MCIU/AEI) under grant 630.

\FloatBarrier
\bibliographystyle{JHEP}
\small \bibliography{bibliography}

\providecommand{\href}[2]{#2}\begingroup\raggedright\begin{thebibliography}{10}

\bibitem{astri22_general}
S.~{Scuderi}, A.~{Giuliani}, G.~{Pareschi}, G.~{Tosti}, O.~{Catalano}, E.~{Amato} et~al., \emph{{The ASTRI Mini-Array of Cherenkov telescopes at the Observatorio del Teide}}, \href{https://doi.org/10.1016/j.jheap.2022.05.001}{\emph{Journal of High Energy Astrophysics} {\bfseries 35} (2022) 52} [\href{https://arxiv.org/abs/2208.04571}{{\ttfamily 2208.04571}}].

\bibitem{pareschi25}
G.~{Pareschi}, \emph{{Status of the ASTRI Mini-Array Gamma-Ray Experiment}},  in \emph{39th International Cosmic Ray Conference (ICRC2025, these proceedings)}, vol.~39 of \emph{International Cosmic Ray Conference}, July, 2025.

\bibitem{vassiliev07}
V.~{Vassiliev}, S.~{Fegan} and P.~{Brousseau}, \emph{{Wide field aplanatic two-mirror telescopes for ground-based {\ensuremath{\gamma}}-ray astronomy}}, \href{https://doi.org/10.1016/j.astropartphys.2007.04.002}{\emph{Astroparticle Physics} {\bfseries 28} (2007) 10} [\href{https://arxiv.org/abs/astro-ph/0612718}{{\ttfamily astro-ph/0612718}}].

\bibitem{sironi17}
G.~{Sironi}, \emph{{Aplanatic telescopes based on Schwarzschild optical configuration: from grazing incidence Wolter-like x-ray optics to Cherenkov two-mirror normal incidence telescopes}},  in \emph{Society of Photo-Optical Instrumentation Engineers (SPIE) Conference Series}, S.L.~{O'Dell} and G.~{Pareschi}, eds., vol.~10399 of \emph{Society of Photo-Optical Instrumentation Engineers (SPIE) Conference Series}, p.~1039903, Sept., 2017, \href{https://doi.org/10.1117/12.2275422}{DOI}.

\bibitem{astri22_science}
S.~{Vercellone}, C.~{Bigongiari}, A.~{Burtovoi}, M.~{Cardillo}, O.~{Catalano}, A.~{Franceschini} et~al., \emph{{ASTRI Mini-Array core science at the Observatorio del Teide}}, \href{https://doi.org/10.1016/j.jheap.2022.05.005}{\emph{Journal of High Energy Astrophysics} {\bfseries 35} (2022) 1} [\href{https://arxiv.org/abs/2208.03177}{{\ttfamily 2208.03177}}].

\bibitem{segreto19}
A.~{Segreto}, O.~{Catalano}, M.C.~{Maccarone}, T.~{Mineo} and A.~{La Barbera}, \emph{{Calibration and monitoring of the ASTRI-Horn telescope by using the night-sky background measured by the photon-statistics (``variance'') method}},  in \emph{36th International Cosmic Ray Conference (ICRC2019)}, vol.~36 of \emph{International Cosmic Ray Conference}, p.~791, July, 2019, \href{https://doi.org/10.22323/1.358.0791}{DOI} [\href{https://arxiv.org/abs/1909.08750}{{\ttfamily 1909.08750}}].

\bibitem{fomin94}
V.P.~{Fomin}, S.~{Fennell}, R.C.~{Lamb}, D.A.~{Lewis}, M.~{Punch} and T.C.~{Weekes}, \emph{{New methods of atmospheric Cherenkov imaging for gamma-ray astronomy. II. The differential position method}}, \href{https://doi.org/10.1016/0927-6505(94)90037-X}{\emph{Astroparticle Physics} {\bfseries 2} (1994) 151}.

\bibitem{hillas85}
A.M.~{Hillas}, \emph{{Cerenkov Light Images of EAS Produced by Primary Gamma Rays and by Nuclei}},  in \emph{19th International Cosmic Ray Conference (ICRC19), Volume 3}, F.C.~{Jones}, ed., vol.~3 of \emph{International Cosmic Ray Conference}, p.~445, Aug., 1985.

\bibitem{crestan25}
S.~{Crestan} and S.~{Lombardi}, \emph{{ASTRI-1: Early data and performance highlights}},  in \emph{39th International Cosmic Ray Conference (ICRC2025, these proceedings)}, vol.~39 of \emph{International Cosmic Ray Conference}, July, 2025.

\bibitem{li83}
T.P.~{Li} and Y.Q.~{Ma}, \emph{{Analysis methods for results in gamma-ray astronomy.}}, \href{https://doi.org/10.1086/161295}{\emph{Astrophysical Journal} {\bfseries 272} (1983) 317}.

\end{thebibliography}\endgroup

\end{document}